\documentclass[pra,twocolumn,showpacs,superscriptaddress]{revtex4-1}
\usepackage{epsfig,graphicx,amssymb,color,bm}
\usepackage{amsmath,float}
\usepackage[colorlinks, linkcolor=blue, urlcolor=blue, citecolor=blue, unicode=true]{hyperref}
\usepackage{setspace}
\usepackage{url}
\usepackage{amsfonts}
\usepackage{latexsym,amssymb}
\usepackage{amsmath, amsbsy}
\usepackage{amsopn, amstext}
\usepackage{graphicx, color, epstopdf}
\usepackage{threeparttable}

\newcommand{\eref}[1]{Eq.~(\ref{#1})}
\newcommand{\esref}[1]{Eqs.~(\ref{#1})}
\newcommand{\fref}[1]{Fig.~\ref{#1}}
\newcommand{\frefs}[1]{Figs.~\ref{#1}}

\begin{document}

\title{Nonlocal nonreciprocal optomechanical circulator\thanks{Project supported by the National Natural Science Foundation of China (Grant Nos. 12061023, 12074206, 11704026, 11704205, 11704042, 11847128).
This work is also sponsored by K.C. Wong Magna Fund in Ningbo University, China}}

% Title should be concise; avoid abbreviations if possible; and not begin with `A', `An', `The', or `Study on'.

\author{Ji-Hui Zheng}
\affiliation{School of Mathematical Sciences, Guizhou Normal University, Guiyang 550025, China}
\affiliation{School of Physical Science and Technology, Ningbo University, Ningbo 315211, China}
\author{Rui Peng}
\affiliation{School of Physics, Dalian University of Technology, Dalian 116024, China}
\author{Jiong Cheng}
\affiliation{School of Physical Science and Technology, Ningbo University, Ningbo 315211, China}
\author{Jing An}\thanks{aj154@163.com}
\affiliation{School of Mathematical Sciences, Guizhou Normal University, Guiyang 550025, China}
\author{Wen-Zhao Zhang}\thanks{zhangwenzhao@nbu.edu.cn}
\affiliation{School of Physical Science and Technology, Ningbo University, Ningbo 315211, China}

\begin{abstract}
A nonlocal circulator protocol is proposed in hybrid optomechanical system.
By analogy with quantum communication, using the input-output relationship, we establish the quantum channel between two optical modes with long-range.
The three body nonlocal interaction between the cavity and the two oscillators is obtained by eliminating the optomechanical cavity mode and verifying the Bell-CHSH inequality of continuous variables.
By introducing the phase accumulation between cyclic interactions, the unidirectional transmission of quantum state between optical mode and two mechanical modes are achieved.
The results show that nonreciprocal transmissions are achieved as long as the accumulated phase reaches a certain value.
 In addition, the effective interaction parameters in our system are amplified, which reduces the difficulty of the implementation of our protocol.
Our research can provide potential applications for nonlocal manipulation and transmission control of quantum platforms.
\end{abstract}
\maketitle

\section{Introduction}
Nonreciprocal quantum state transmission has emerged as an indispensable tool for the important applications in quantum information processing such as optical diode \cite{PhysRevLett.110.093901,PhysRevA.90.023849,PhysRevA.91.063836}, noise-free sensing \cite{PhysRevA.99.063811}, unidirectional amplifier \cite{PhysRevA.100.043835}, and nonreciprocal phase shifter \cite{PhysRevApplied.13.044040}.
A nonreciprocal response can be generated by direct broken the time-reversal symmetry in the transmission.
Many protocols have been proposed in both theoretical and experimental perspective to achieve this propose.
In the electromagnetic domain, special magnetic effects are explored and widely used for nonreciprocal response \cite{PhysRevLett.126.177401,PhysRevB.102.134417,PhysRevLett.123.077401}.
In the optical domain, directional transmission of light has been achieved with optical nonlinearities or dynamically modulated media \cite{PhysRevLett.125.123901,PhysRevApplied.15.044041}.
In the atomic domain, nonreciprocal cyclic transition was proposed in a multi-level atomic system or topological cold-atom systems \cite{PhysRevLett.123.033902,PhysRevLett.123.180402,PhysRevLett.124.250402}.
Nonreciprocal effect are exported in many typical quantum systems.
Corresponding applications are reported to realized quantum device based on  nonreciprocal response.

Hybrid quantum operations are usually used to optimize or implement multiple quantum processing.
Recently, nonreciprocal transmission in optomechanical system has attracted more and more attention for its potential applications in hybrid quantum systems and hybrid quantum network \cite{PhysRevA.102.011502,PhysRevLett.125.023603,PhysRevLett.120.023601}.
The optomechanical system was referred to as a natural  bridge between photons and phonons \cite{RevModPhys.86.1391}.
Thus, the controlling and  transmission between solid state and flying state are subsequently realized in such system.
For example, nonreciprocal phonon transport based on arrays of optomechanical microtoroids \cite{PhysRevB.101.085108}, nonreciprocal conversion between microwave and optical photons in electro-optomechanical systems \cite{PhysRevA.93.023827} and unidirectional amplification in optical gain optomechanical systems \cite{PhysRevA.100.043835}.
Up to now, most of the studies are limited in the localized quantum effect to achieve nonreciprocal properties.
Nonlocality, as a distinguishing feature of quantum mechanics from classical physics, remains unexplored when studying the quantum nonreciprocal effect in optomechanical systems.
Therefore, it is necessary to find that if the nonlocal manipulation of nonreciprocal transmissions could be achieved in optomechanical systems.

Remote transmission or distribution of optical quantum state has been well studied in optical fiber and free space \cite{science.356.1140}, such as satellite-based entanglement distribution with long-distance \cite{PhysRevLett.123.150402}, nonlocal quantum state transmission in optical fiber \cite{PhysRevA.90.012324}, etc.
The construction of quantum nonlocality based on this technology has been widely reported.
Thus, it is possible for us to realize nonlocal interaction in optomechanical systems by utilizing the remote quantum technique.
In this work, a scheme to achieve nonlocal nonreciprocal transmission is proposed in composite optomechanical systems.
A nonlocal interaction is constructed between the remote optical and two mechanical modes.
The time-reversal symmetry breaking is achieved through the nonlocal interaction of three bodies, so as to realize the nonlocal one-way quantum state transmission and control between optical and mechanical modes.
This paper is organized as follows.
In Sec.~2, we analyze the model and Hamiltonian according to the dynamic properties of the system.
By using the linearization approximation and adiabatic elimination method, we obtain the nonlocal effective Hamiltonian which is used to describe the long-range interaction between optical mode and two mechanical modes.
We then study the optomechanical circulator behavior in Sec.~3.
Eventually, discussions and conclusions are given in Sec.~4.

\section{Model and Hamiltonian} \label{sec2}
\begin{figure}
  \centering
  \includegraphics[width=8.5cm]{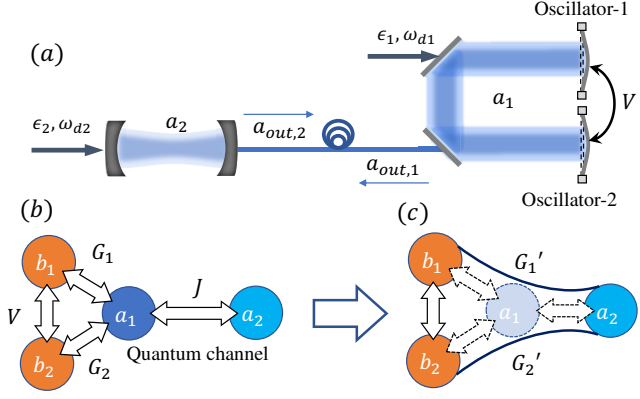}\\
  \caption{(a) The schematic diagram of a nonreciprocal optomechanical circulator system.
The monochromatic cavity field is coupled with two movable mirrors.
Lossless fiber (or free space) is used to construct quantum channel between optomechanical cavity and empty cavity.
(b) The interaction structure diagram of the system and the nonlocal interaction diagram we expected.}\label{fig1}
\end{figure}

\begin{figure}
  \centering
  \includegraphics[width=5.25cm]{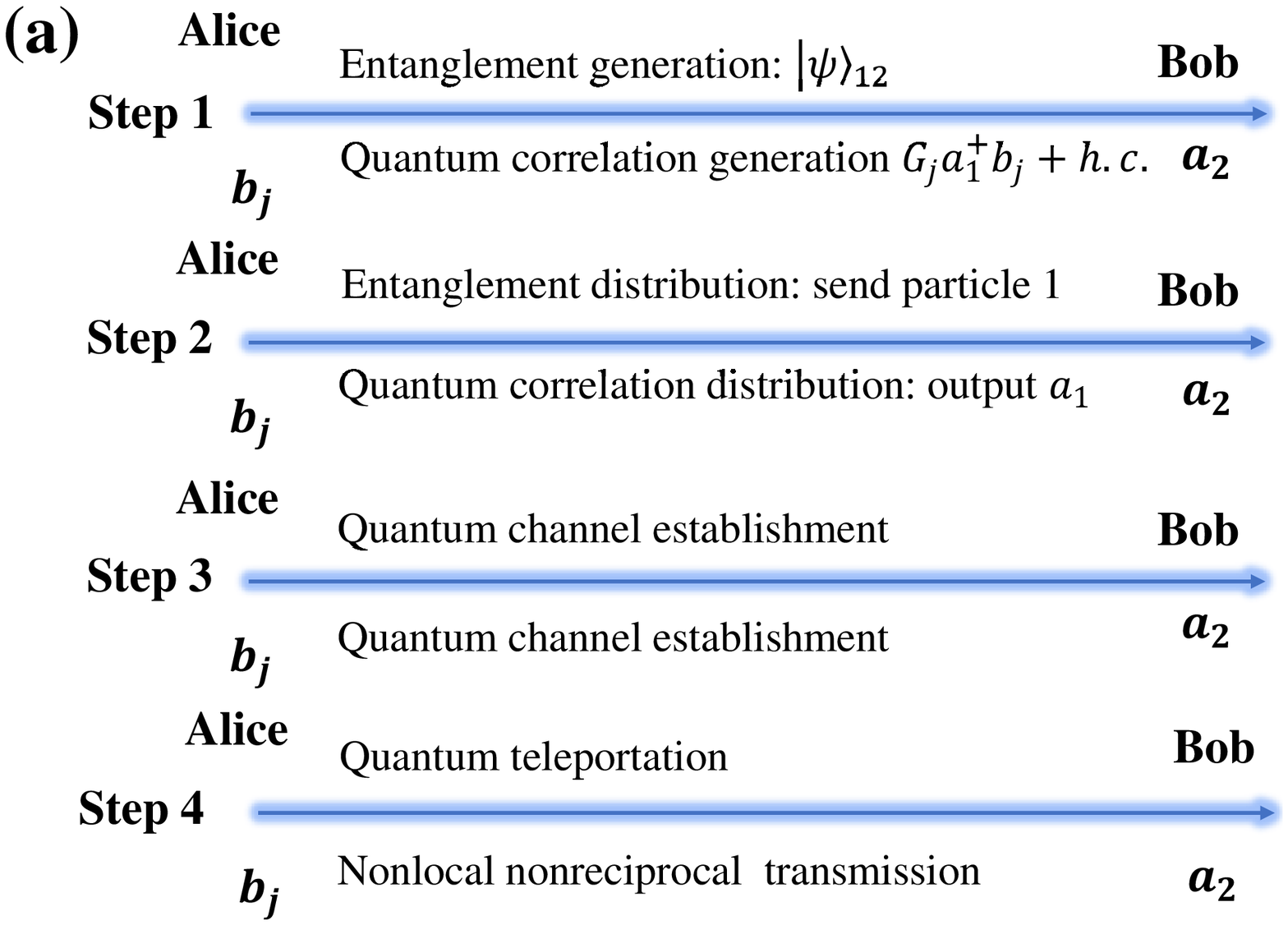}
  \includegraphics[width=3.25cm]{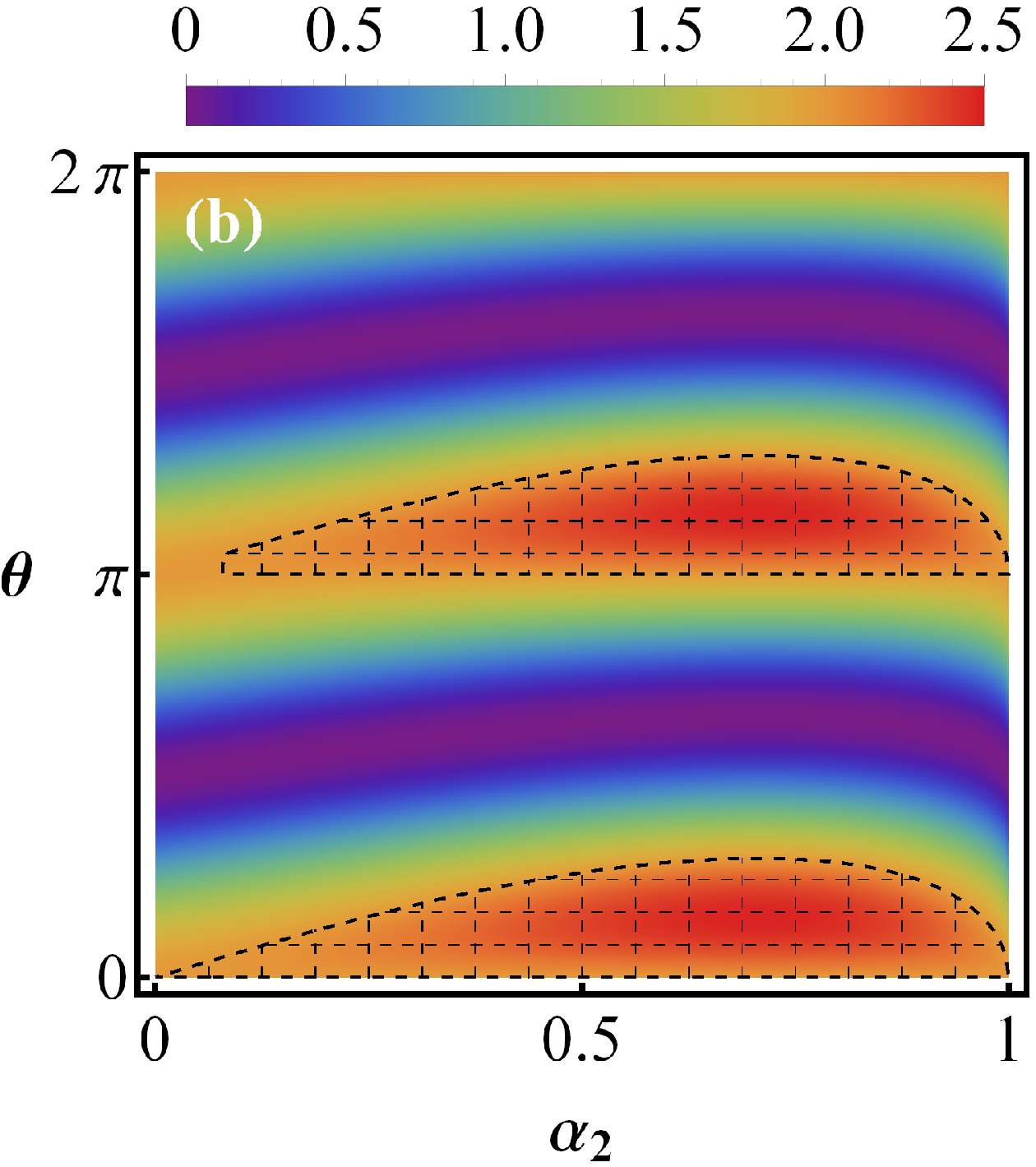}\\
  \caption{(a) The step analogy between quantum teleportation and nonlocal nonreciprocal transmission,  where the lower part of the line represents our protocol.
  (b) The average value of Bell-CHSH operator for continuous variable (CV) systems ($|\langle \mathcal{B}_{CHSH} \rangle| $) as a function of angle $\theta$ and superposition coefficient $\alpha_2$.
  The black grid region represents the violation of Bell-CHSH inequality, that is, the parameters region with $|\langle \mathcal{B}_{CHSH} \rangle|>2$.}\label{figadd}
\end{figure}

As shown in \fref{fig1}, the system consists of three parts: a composite optomechanical system, a empty cavity and  a long-range lossless optical fiber (or free space).
The long-range lossless optical fiber is used to transmits the output photons from cavity 1 and 2.
The Hamiltonian of the system is written as
\begin{eqnarray}
\hat{H}&=&\hat{H}_{sys}+\hat{H}_d,\\
\hat{H}_{sys}&=&\sum_{j=1}^{2}\omega_{cj}\hat{a}_j^{\dag}\hat{a}_j+\omega_{mj}
\hat{b}_j^{\dag}\hat{b}_j\nonumber\\
&&+g_j\hat{a}_1^{\dag}\hat{a}_1(\hat{b}_j^{\dag}+\hat{b}_j)+V\hat{b}_j^{\dag}\hat{b}_{3-j},\\
\hat{H}_d&=&\sum_{j=1}^2\varepsilon_j(\hat{a}_j^{\dag}e^{-i\omega_{dj}t}+\hat{a}_je^{i\omega_{dj}t}),
\end{eqnarray}
where $\hat{a}_j (\hat{a}_j^{\dag}) $ is the annihilation (creation) operator of the cavity mode $\hat{a}_j$ with resonance frequency $\omega_{cj}$, $\hat{b}_j (\hat{b}_j^{\dag}) $ is the annihilation (creation) operator of the mechanical mode $\hat{b}_j $ with resonance frequency $\omega_{mj}$, and $g_j$ is the single-photon optomechanical coupling strength between the cavity mode $\hat{a}_1$ and the mechanical mode $\hat{b}_j $.
The fourth term denotes the interaction between the mechanical resonator $\hat{b}_1$ and $\hat{b}_2$ with coupling strength  $V$.
${H}_d$ is the external driving field with the intensity of $\varepsilon_j$, and the frequency $\omega_{dj}$.
Under the rotating reference frame at the frequency $\omega_{dj}$ of the driving field frequency.
The quantum Langevin equations of the system are written as
\begin{subequations}
\begin{eqnarray}
 \dot{\hat{a}}_1 &=& -[i\Delta_{c1}+\sum_{j=1}^2i g_j(\hat{b}_j^{\dag}+\hat{b}_j)+\frac{\kappa_1}{2}]\hat{a}_1\nonumber\\
&&-i\varepsilon_1+\sqrt{\kappa}_1\hat{a}_{in,1}', \label{eqn4}\\
 \dot{\hat{a}}_2 &=& -(i\Delta_{c2}+\frac{\kappa_2}{2})\hat{a}_2-i\varepsilon_2+\sqrt{\kappa}_2\hat{a}_{in,2}', \label{eqn5}\\
\dot{\hat{b}}_j&=&-(i\omega_{mj}+\frac{\gamma_j}{2})\hat{b}_j-ig_j\hat{a}_1^{\dag}\hat{a}_1\nonumber\\&&-iV\hat{b}_{3-j}s+\sqrt{\gamma_j}\hat{b}_{in,j},
\end{eqnarray}
\end{subequations}
where $\Delta_{cj}=\omega_{cj}-\omega_{dj}$ is the detuning between the driving field and the cavity, $\kappa_j$ denotes the cavity damping rate, $\gamma_j$ represents the damping rate of the mechanical mode.
$\hat{b}_{in,j}$ is noise operator of $j$-th mechanical oscillator.
The term $\hat{a}'_{in,j}$ represents the input operator of $j$-th cavity, which consists of the output of the other cavity and the noise input $\hat{a}_{in,j}$.
According to the input-output formalism \cite{PhysRevA.31.3761} for optical cavity, i.e. $\hat{O}_{out}=\sqrt{\kappa}\hat{O}-\hat{O}_{in}$, one can obtain that \cite{PhysRevA.31.3761},
\begin{equation}
 \hat{a}_{in,j}'=(\sqrt{\kappa_{3-j}}\hat{a}_{3-j}-\hat{a}_{in,3-j}) e^{-i\phi}+\hat{a}_{in,j},
\end{equation}
where $\phi$ is the phase delay caused by the remote fiber, which can be adjusted by the phase retarder.
For convenience, we set $\phi=\pi/2$.
By substituting the expressions of $\hat{a}_{in,j}'$ into Eqs.~\eqref{eqn4}-\eqref{eqn5}, the Langevin equations of the system are rewrite as
\begin{subequations}
\begin{eqnarray}
\dot{\hat{a}}_1&=& -[i\Delta_{c1}+\sum_{j=1}^2i g_j(\hat{b}_j^{\dag}+\hat{b}_j)+\frac{\kappa_1}{2}]\hat{a}_1\nonumber\\
&&-i\varepsilon_1-i J\hat{a}_2+\sqrt{\kappa}_1\hat{A}_{in,1},\\
\dot{\hat{a}}_2 &=&  -(i\Delta_{c2}+\frac{\kappa_2}{2})\hat{a}_2-i\varepsilon_2-iJ\hat{a}_1+\sqrt{\kappa}_2\hat{A}_{in,2},\\
\dot{\hat{b}}_j&=&-(i\omega_{mj}+\frac{\gamma_j}{2})\hat{b}_j-ig_j\hat{a}_1^{\dag}\hat{a}_1\nonumber\\&&-iV\hat{b}_{3-j}+\sqrt{\gamma_j}\hat{b}_{in,j},
\end{eqnarray}
\end{subequations}
where $J=\sqrt{{\kappa}_1{\kappa}_2}$, denotes the remote interaction strength between optical modes $1$ and $2$.
The corresponding input noise operators of the cavity models are $\hat{A}_{in,j}=\hat{a}_{in,j}+i\hat{a}_{in,3-j}$.
According to the original Hamiltonian, there is no direct interaction between $a_1$ and $a_2$.
Thus, in our model, $J$ is the key to construct nonlocal interaction.
In fact, this interaction results from the correlation between the intracavitary field and the output field of cavities, and the transmission  process between $a_1$ and $a_2$ is similar to the entanglement distribution in quantum teleportation.
The quantum channel between  $a_1$ and $a_2$ can be established through this remote correlation.
Once the quantum channel is formed, we can achieve the nonlocal interaction between $a_1$ and $a_2$ just like quantum teleportation.
The step analogy between quantum teleportation and our scheme is displayed in \fref{figadd} (a).
In our model, this is the resources to realize nonlocal nonreciprocal transmission.
The process of information output through $\hat{b}_j$($ \hat{a}_2$) mode and the dissipation throughout the system will consume this correlation.
When this kind of resource is exhausted (when the correlation between photons disappears), we need to redistribute the nonlocal correlation photons.
Thus, our protocol needs to have a correlation distribution process similar to entanglement distribution to maintain nonlocal and nonreciprocal transmission.

Under strong driving and weak coupling condition, the so called `` linearized approximation" \cite{RevModPhys.86.1391} is used to decomposed the operator into average value and its fluctuation term, i.e.,
 $\hat{a}_j=\alpha_j+\delta\hat{a}_j$, $\hat{b}_j=\beta_j+\delta\hat{b}_j$, where $\alpha_j\equiv \langle\hat{a}_j \rangle$ is the mean values of $\hat{a}_j$, $\beta_j\equiv \langle\hat{b}_j \rangle $ is the mean values of $\hat{b}_j$ respectively.
After choose the suitable sideband with $\Delta_{c1}' \approx \omega_{mj}$.
The linearized Langevin equations of the system under the rotating-wave approximation are then governed by,
\begin{subequations}
\begin{eqnarray}\label{effle}
\delta\dot{\hat{a}}_1&=&-(i\Delta_{c1}'+\frac{\kappa_1}{2})\delta\hat{a}_1-i\sum_{j=1}^2G_j\delta\hat{b}_j\nonumber\\&&-iJ\delta\hat{a}_2+\sqrt{\kappa_1}\hat{A}_{in,1},\label{qpart1}\\
\delta\dot{\hat{a}}_2&=& -(i\Delta_{c2}+\frac{\kappa_2}{2})\delta\hat{a}_2-iJ\delta\hat{a}_1+\sqrt{\kappa_2}\hat{A}_{in,2}, \\
\delta\dot{\hat{b}}_j&=&-(i\omega_{mj}+\frac{\gamma_j}{2})\delta\hat{b}_j-iG_j^*\delta\hat{a}_1\nonumber\\&&-iV\delta\hat{b}_{3-j}+\sqrt{\gamma_j}\hat{b}_{in,_j},\label{qpart2}
\end{eqnarray}
\end{subequations}
where $\Delta_{c1}'=\Delta_{c_1}+\sum_{j=1,2}g_j(\beta_j+\beta_j^*)$ represents the effective detuning modulated by the mechanical average position.
$G_j=g_j\alpha_1$ are the linearized optomechanical coupling strength.

The interaction relationship diagram in the system is displayed in \fref{fig1} (b).
We aim to eliminate the optical mode $\delta \hat{a}_1$, so as to realize the long-range interaction between optical mode $\delta \hat{a}_2$ and mechanical modes, i.e. \fref{fig1} (c).
By setting the strong dissipation of the optomechanical cavity, we can regard that the optical model $a_1$ is fast dissipative in the evolution of the system.
Thus, the rapid dissipation mode $a_1$ can be eliminated under the condition $\kappa_1\gg\{\gamma_1, \gamma_2, \kappa_2\}$.
This method as been reported and confirmed in Refs.~\cite{PhysRevLett.110.153606,OE.25.10779}.
In this case, the nonlocal interaction between the optical mode $\delta \hat{a}_2$ and the mechanical modes are obtained.
This effective three-mode optomechanical system can be used as a three-port circulator \cite{PhysRevA.91.053854} for one optical mode and two mechanical modes.
The corresponding effective Langevin equations are (details are shown in appendix A),
\begin{align}
\delta\dot{\hat{a}}_2=&-(i\Delta_{eff}+\frac{\kappa_{eff}}{2})\delta\hat{a}_2+i\sum_j G_j'\delta\hat{b}_j+\sqrt{\kappa_{eff}}\hat{A}_{in},\nonumber\\
\delta\dot{\hat{b}}_j=&-(i\omega_{effj}+\frac{\gamma_{effj}}{2})\delta\hat{b}_j\nonumber\\&+iG_j''\delta\hat{a}_2+iV_j\delta\hat{b}_{3-j}+\sqrt{\gamma_{effj}}\hat{B}_{in,j} ,\label{eqeffl}
\end{align}
where $\sqrt{\kappa_{eff}}\hat{A}_{in}=-iJ\hat{a}_{in}+\sqrt{\kappa_2}\hat{A}_{in,2}$, $\sqrt{\gamma_{effj}}\hat{B}_{in,j}=-iG_j^*\hat{a}_{in}+\sqrt{\gamma_j}\hat{b}_{in,j}$.
The effective parameters are expressed as,
\begin{eqnarray}\label{peff}
\Delta_{eff}&=&\Delta_{c2}-\xi_c(\Delta_{c1}'-\Delta_{c2}),\nonumber\\
\omega_{effj}&=&\omega_{mj}-\xi_{mj}(\Delta_{c1}'-\omega_{mj}),\nonumber\\
\kappa_{eff}&=&\kappa_2+\xi_1(\kappa_1-\kappa_2),\nonumber\\
\gamma_{effj}&=&\gamma_j+\xi_{mj}(\kappa_1-\gamma_j),\nonumber\\
G_j'&=& J G_j \xi_j,\nonumber\\
G_j'' &=& J G_j^* \xi,\nonumber\\
V_j &=& G_j^* G_{3-j} \xi_{3-j}-V,
\end{eqnarray}
the corresponding effective correction factors are
\begin{eqnarray*}
\xi_c &=& \frac{J^2}{(\Delta_{c1}'-\Delta_{c2})^2+\frac{(\kappa_1-\kappa_2)^2}{4}},\\
\xi_{mj} &=& \frac{|G_j|^2}{(\Delta_{c1}'-\omega_{mj})^2+\frac{(\kappa_1-\gamma_j)^2}{4}},\\
\xi_j &=& \frac{(\Delta_{c1}'-\omega_{mj})+\frac{i(\kappa_1-\gamma_j)}{2}}{(\Delta_{c1}'-\omega_{mj})^2+\frac{(\kappa_1-\gamma_j)^2}{4}},\\
\xi &=& \frac{(\Delta_{c1}'-\Delta_{c2})+\frac{i(\kappa_1-\kappa_2)}{2}}{(\Delta_{c1}'-\Delta_{c2})^2+\frac{(\kappa_1-\kappa_2)^2}{4}}.
\end{eqnarray*}
It is obvious that, the effective parameters strongly depends on the above correction coefficient.
Moreover, as long as we select the appropriate sideband, the amplified effective parameters can be obtained.
In addition, under appropriate parameter conditions, we can write back the effective Hamiltonian according to the effective Langevin equations (see appendix A for details discussion).
This effective dynamics or Hamiltonian can provide us with a powerful means to further confirm that our scheme is nonlocal, the Bell-CHSH inequality for continuous variable systems is tested \cite{PhysRevLett.88.040406,PhysRevLett.68.3259}.
Under the sideband condition, the average value of $\mathcal{B}_{CHSH}$ on Fock state is expressed as (details see appendix B),
\begin{eqnarray}
\langle\mathcal{B}_{CHSH} \rangle &=& 2 ( \cos \theta+\sin\theta  \alpha_2 \sqrt{1-|\alpha_2|^2})^2,
\end{eqnarray}
where $\theta_a=\theta_b=\theta$ denotes the angles of unit vectors used to verify the inequality, 
$\alpha_2$ represents the superposition coefficients of the initial uncoupled states in the final state after evolution.
When $G_j'=0$, the coefficients $\alpha_2=1$, the maximum value of $\langle\mathcal{B}_{CHSH} \rangle$ is $2$, which does not violate Bell-CHSH inequality, the system embodies locality.
The nonlocality of our system will be confirmed as long as the inequality $|\langle\mathcal{B}_{CHSH} \rangle |\leq2$ is violated.
In black grid region in \fref{figadd}(b), the results show that $|\langle \mathcal{B}_{CHSH} \rangle|>2$.
The violation of the inequality here depends on $G_j'$ in \eref{peff}, that is, the remote interaction strength mentioned in our previous analysis.

\begin{figure}[htb]
  \centering
  \includegraphics[width=8.5cm]{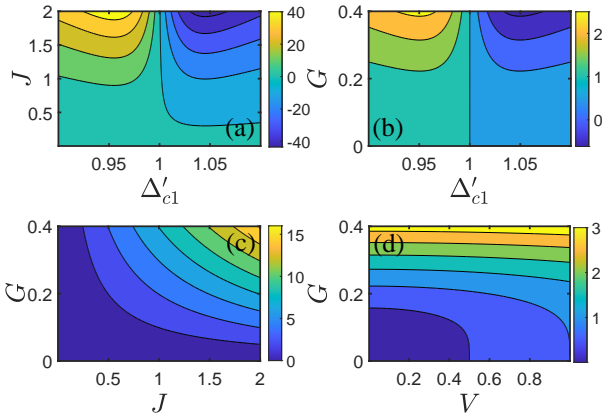}\\
  \caption{The effective parameters $\Delta_{eff},\omega_{eff},G_j',V_j$ as a function of the original parameters $J,\Delta_c',G,V$ (unit of $\omega_m^{-1}$).
  Other parameters are $\gamma=10^{-3}\omega_m$, $\kappa_1=10 \kappa_2=\omega_m/10$, $\Delta_{c2}=\omega_m$ and $\Delta_{c1}'-\Delta_{c2}=10^{-4}\omega_m$.}\label{fig2}
\end{figure}

As shown in \fref{fig2}, the dependence of the effective parameters $\Delta_{eff},\omega_{effj},G_j'$ and $V_j$ on the original parameters are displayed in the sub-figures (a), (b), (c) and (d), respectively.
For the convenience of discussion, we set $\omega_{mj}=\omega_m$, $\gamma_j=\gamma$ and $|G_j|=G$.
\fref{fig2}(a) shows the effective detuning as the function of $\Delta_c$ and $J$.
It's obvious that, $\Delta_{eff}$ can be adjusted from $-40 \omega_{m}$ to $40\omega_{m}$ with small original parameters change i.e. $J/\omega_{m}\in [0,2]$ and $\Delta_{c1}'/\omega_{m}\in [0.9,1.1]$.
Other effective parameters $\omega_{effj},G_j'$ and $V_j$ can also be adjusted in a large range with the small change of the original parameters $\{ G, J, \Delta_{c1}', V \}$.
The corresponding results are shown in the sub-figures (b), (c) and (d), respectively.

\begin{figure}[htb]
  \centering
  \includegraphics[width=8.5cm]{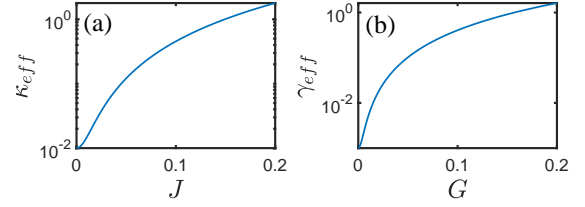}\\
  \caption{The effective dissipation coefficient $\kappa_{eff},\gamma_{eff}$ as a function of the original parameters $J,G$.
  Other parameters are the same with \fref{fig2}.} \label{fig3}
\end{figure}

According to the expression of \esref{peff}, the effective dissipation coefficient $\kappa_{eff}$ and $\gamma_{eff}$ are modulated by the original parameters.
Therefore, when we use the original parameters to enhance the coupling rate, the effect of dissipation should also be taken into account.
The relationship between effective dissipation and coupling coefficient are shown in \fref{fig3}.
The effective dissipation increases with the increase of coupling coefficient $J$ and $G$.
In the following discussion, we will consider both the enhancement of effective coupling and the change of effective dissipation from the original parameters.

\section{Nonlocal circulator}\label{sec3}
In this section, we mainly studies how to realize the nonreciprocal transmission of photon-phonon in the three mode nonlocal optomechanical system.
Perfect nonreciprocal transmission means that signals can be transmitted completely from one side of the system to the other while blocking reverse transmission.
To achieve this nonreciprocal transmission, we characterize the input-output relations of such a device and shows the adjustability of the system.
For simplicity, the above Heisenberg-Langevin equations \eqref{eqeffl} are written in compact form,
\begin{align}
& \dot{\hat{Q}}=-M\hat{Q}+L\hat{Q}_{in} \label{eqn20},
\end{align}
where the fluctuation vector is $\hat{Q}=(\delta\hat{a}_2,\delta\hat{b}_1,\delta\hat{b}_2)^T$, the input field vector is $\hat{Q}_{in}=(\hat{A}_{in},\hat{B}_{in,1},\hat{B}_{in,2})^T$, and the damping matrix is $L=diag(\sqrt{\kappa_{eff}},\sqrt{\gamma_{eff1}},\sqrt{\gamma_{eff2}})$.
Further, $M$ denotes the coefficient matrix, namely,

\begin{equation}
\mathbf{M}=\left(
\begin{array}{cccc}
\frac{\kappa_{eff}}{2}+i\Delta_{eff} & -iG_1' & -iG_2'\\
-iG_1'' & \frac{\gamma_{eff1}}{2}+i\omega_{eff1} & -iV_1\\
-iG_2'' & -iV_2 & \frac{\gamma_{eff2}}{2}+i\omega_{eff2}\\
\end{array}
\right).
\end{equation}

 Introducing the Fourier transformation on \eref{eqn20} \cite{PhysRevA.91.053854,PhysRevA.96.053853}, and using the input-output formalism \cite{PhysRevA.31.3761}.
 The output field vector in the frequency domain is
\begin{align}
&\hat{Q}_{out}(\omega)=\Gamma(\omega)\hat{Q}_{in}(\omega),
\end{align}
the transmission evolution matrix is $\Gamma(\omega)=L[M-i\omega I]^{-1}L-I$.
According to the definition of spectrum, the relationship between input and output spectra is expressed as
\begin{eqnarray}
S_{out}(\omega)&=& T(\omega)S_{in}(\omega)+S_{vac}(\omega),
\end{eqnarray}
where $S_{vac}$ denotes the output spectrum contributing from the input vacuum field.
Under rotating wave approximation, $S_{vac}$ can be ignored.
The corresponding scattering matrix $T$ then be expressed as
\begin{equation}
\mathbf{T}(\omega)=\left(
\begin{array}{cccc}
T_{\hat{a}_2\hat{a}_2}(\omega) & T_{\hat{a}_2\hat{b}_1}(\omega) & T_{\hat{a}_2\hat{b}_2}(\omega)\\
T_{\hat{b}_1\hat{a}_2}(\omega) & T_{\hat{b}_1\hat{b}_1}(\omega) & T_{\hat{b}_1\hat{b}_2}(\omega)\\
T_{\hat{b}_2\hat{a}_2}(\omega) & T_{\hat{b}_2\hat{b}_1}(\omega) & T_{\hat{b}_2\hat{b}_2}(\omega)\\
\end{array}
\right),
\end{equation}
where
\begin{align}
& T_{\hat{a}_2\hat{a}_2}(\omega)=|\Gamma_{11}(\omega)|^2,\nonumber\\
& T_{\hat{a}_2\hat{b}_1}(\omega)=|\Gamma_{12}(\omega)|^2,\nonumber\\
& T_{\hat{a}_2\hat{b}_2}(\omega)=|\Gamma_{13}(\omega)|^2,\nonumber\\
& T_{\hat{b}_1\hat{a}_2}(\omega)=|\Gamma_{21}(\omega)|^2,\nonumber\\
& T_{\hat{b}_1\hat{b}_1}(\omega)=|\Gamma_{22}(\omega)|^2,\nonumber\\
& T_{\hat{b}_1\hat{b}_2}(\omega)=|\Gamma_{23}(\omega)|^2,\nonumber\\
& T_{\hat{b}_2\hat{a}_2}(\omega)=|\Gamma_{31}(\omega)|^2,\nonumber\\
& T_{\hat{b}_2\hat{b}_1}(\omega)=|\Gamma_{32}(\omega)|^2,\nonumber\\
& T_{\hat{b}_2\hat{b}_2}(\omega)=|\Gamma_{33}(\omega)|^2,\nonumber
\end{align}
$T_{x,y}$ denotes the transmission rate from mode $x$ to mode $y$.
In the aspect of experimental simulation, we generally pay more attention to the adjustable parameters in the experiment.
We noticed that, mode $b_1$ and $b_2$ are completely symmetric in the form of dynamics.
Therefore, when considering the influence of parameters, we only need to pay attention to the parameters of one of the modes.

\begin{figure}[htb]
  \centering
  \includegraphics[width=8.5cm]{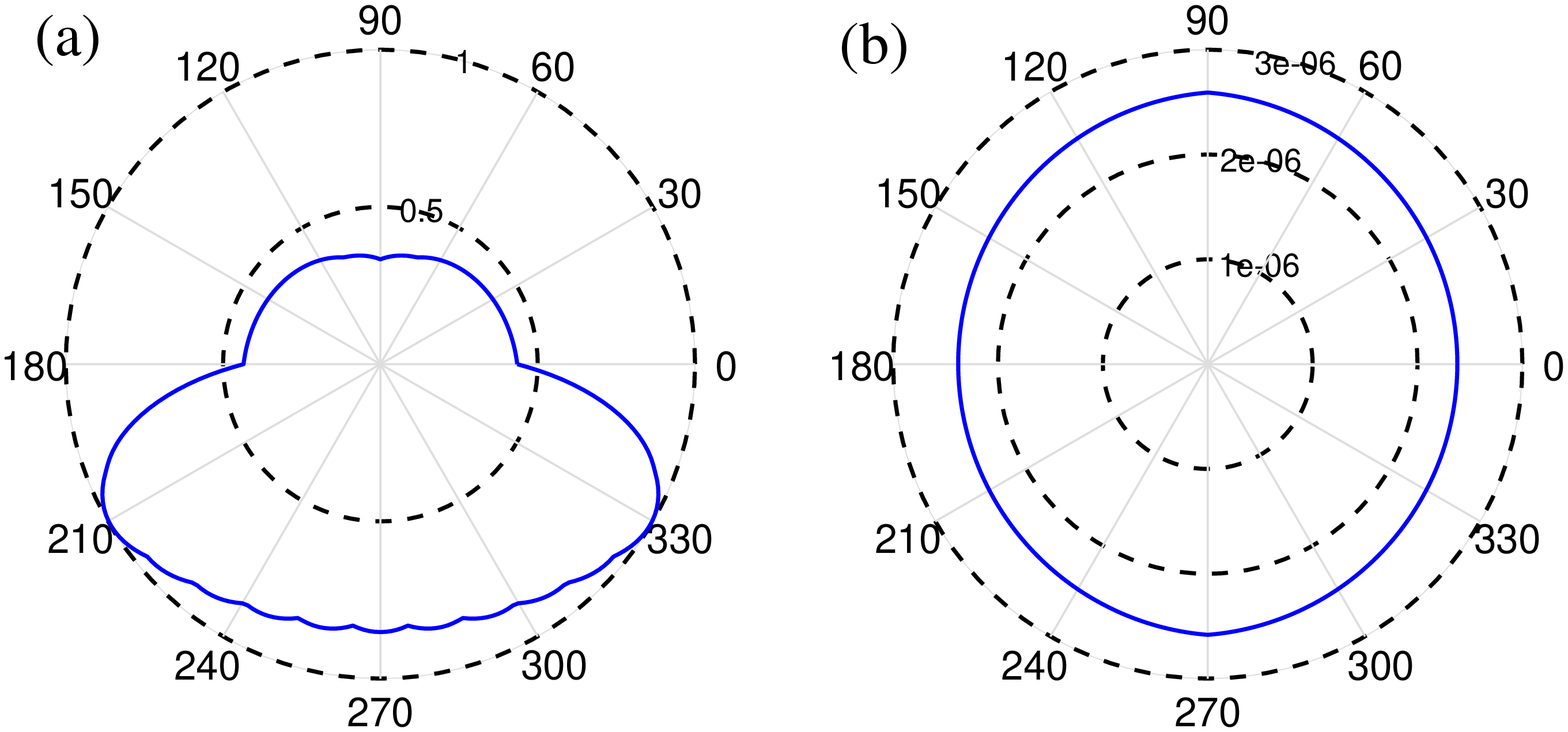}\\
  \caption{The transmission rate $\textbf{Max}\left[ T_{\hat{a}_2\hat{b}_1}(\omega)\right ]$ and $\textbf{Min}\left[ T_{\hat{b}_1\hat{a}_2}(\omega)\right ]$ as a function of the phase factor $\theta_1$(rad).
  Other parameters are the same with \fref{fig2}.}\label{fig4}
\end{figure}

The effective coupling rate $G_j'$ and $V_j$ are complex, without lose of generality, we can set $G_j'=G_{j0} \exp(-i \theta_j)$ and $V_j=V_0 \exp[(-1)^ji \theta_3]$.
The phases of the coupling rate plays an important role in nonreciprocal transmission, which can introduce a symmetry breaking of the system modes coupling and then to achieve nonreciprocal transmission.
The corresponding effective of the phase factor $\theta_1$ on the max value of $T_{\hat{a}_2\hat{b}_1}(\omega)$ and min value of $T_{\hat{b}_1\hat{a}_2}(\omega)$ is displayed in \fref{fig4}.

As shown in \fref{fig4}(a), it's obvious that the phase factor has a significant influence on the maxmal transmission rate $T_{\hat{a}_2\hat{b}_1}(\omega)$.
The maximum value of the transmission rate can reach 1 when the appropriate phase factor is selected.
According to the figure, $\textbf{Max}\left[ T_{\hat{a}_2\hat{b}_1}(\omega)\right ]=1$ are reached around $210^{\circ}$ and $330^{\circ}$.
In \fref{fig4}(b), the influence of phase factor on minimal value of $T_{\hat{b}_1\hat{a}_2}(\omega)$ can be ignored.
For any value of $\theta_1$, $\textbf{Min}\left[ T_{\hat{b}_1\hat{a}_2}(\omega)\right ]$ is closed to zero, that is, phase have not affect on the block of $T_{\hat{b}_1\hat{a}_2}$ transmission.
To achieve nonreciprocity, it is necessary to select appropriate parameters to achieve $T_{\hat{a}_2\hat{b}_1}\approx 1$ and $T_{\hat{b}_1\hat{a}_2}\approx 0$ at the same $\omega$.

\begin{figure}[htb]
  \centering
  \includegraphics[width=9cm]{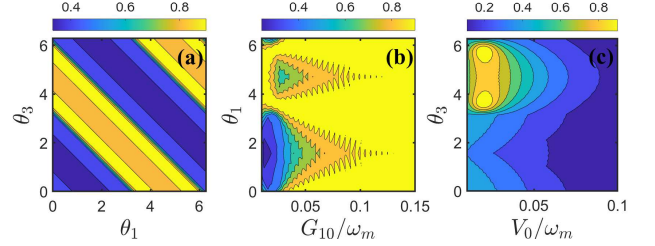}\\
  \caption{The transmission rate $\textbf{Max}\left[ T_{\hat{a}_2\hat{b}_1}(\omega)\right ]$ as a function of the coupling parameters $\theta_1$, $\theta_3$, $G_{10}$ and $V_0$.
  Other parameters are the same with \fref{fig2}.}\label{fig5}
\end{figure}

For further discuss the nonreciprocity effect in our protocol, we show the relationship of transmission rate $\textbf{Max}\left[ T_{\hat{a}_2\hat{b}_1}(\omega)\right ]$ and adjustable parameters: $\theta_1$, $\theta_3$, $G_{10}$ and $V_0$.
The corresponding results are displayed in \fref{fig5}.
As shown in \fref{fig5}(a),
the maximum transmission rate $\theta_1$ and $\theta_3$ on $\textbf{Max}\left[ T_{\hat{a}_2\hat{b}_1}(\omega)\right ]$ increases and decreases periodically with the increase of $\theta_1$ and $\theta_3$.
When the value of $\theta_1$ and $\theta_3$ meet the appropriate matching conditions, that is, the bright yellow region in the figure, the maximum transmission rate is close to 1.
As shown in \fref{fig5}(b), We investigate the influence of $\theta_1$ and $G_{10}$ on the maximum transmission rate.
It is obvious that when $G_{10}$ is larger enough, no matter what the phase factor is, we have $\textbf{Max}\left[ T_{\hat{a}_2\hat{b}_1}(\omega)\right ]=1$.
According to the effective Langevin equations \esref{effle}, $G_{10}$ represents the coupling strength between the optical mode $\hat{a}_2$ and the mechanical mode $\hat{b}_1$, which directly determines the information transmission ability between this two modes.
When $G_{10}\gg \{G_{20},V_0\}$, the coupling between $\hat{a}_2$ and $\hat{b}_1$ is dominant in the system,
under this condition, the system can be approximated as a two-body interaction system.
The interaction of our system shows a symmetrical structure, thus the phase modulation effect is suppressed.
Although the maximum transmission rate $T_{\hat{a}_2\hat{b}_1}$ can be achieved, nonreciprocity effect will disappear.
When the value of $G_{10}$ close to $\{G_{20},V_0\}$.
The system is a three body circulator.
Under this condition, the influence of phase factor is highlighted.
There are obvious extremums of $\theta_1$ around $7/6 \pi$ and $11/6 \pi$.
As shown in \fref{fig5}(c), when $V_{0}$ is larger enough, no matter what the phase factor is, we have $\textbf{Min}\left[ T_{\hat{a}_2\hat{b}_1}(\omega)\right ]\approx 0$.
This is due to the symmetry of the dominant coupling of $\hat{b}_1$ and $\hat{b}_2$.
When $V_{0}\gg \{G_{10},G_{20}\}$, the coupling between $\hat{a}_2$ and $\hat{b}_1$ can be ignored.
When the value of $V_{0}$ close to $\{G_{10},G_{20}\}$.
Under this condition, the influence of phase factor is highlighted.
There are obvious extremums of $\theta_3$ around $7/6 \pi$ and $11/6 \pi$.
Therefore, choosing appropriate parameters can realize and control nonreciprocity.
To realize the three body circulator, the chose of the coupling rate $V_0$ and $G_{j0}$ should be closed.
And the phase of the coupling rate also needs to meet the matching conditions.

\begin{figure}[htb]
  \centering
  \includegraphics[width=8.5cm]{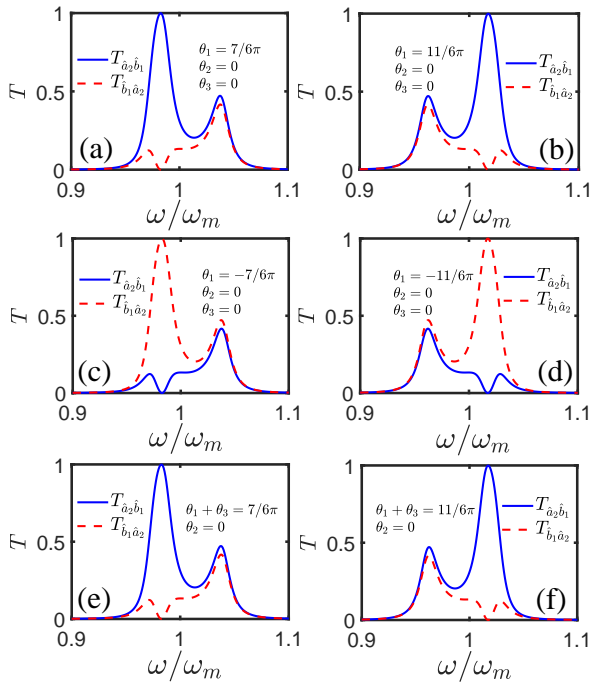}\\
  \caption{The transmission spectrum with different phase factors.
  Other parameters are the same with \fref{fig2}.}\label{fig6}
\end{figure}

In stand three body circulator, the signal is transferred from one mode to another in either a counterclockwise ($\hat{a}_2 \rightarrow \hat{b}_1 \rightarrow\hat{b}_2 \rightarrow \hat{a}_2$ ) or clockwise direction, depending on the relative phase \cite{PhysRevA.91.053854}.
In our model, according to the conclusion in \fref{fig4}, this relative phase $\theta=7 \pi/6$ or $\theta=11\pi/6 $.
As shown in \frefs{fig6}(a)-(f), under the condition of phase matching, our system shows obvious nonreciprocity.
This is consistent with the conclusions of our previous analysis.
In \fref{fig6}(e) and (f), we show that the selection of each phase can be arbitrary, as long as the phase accumulation on the circulator satisfies the condition.
In the above process of nonreciprocal transmission, the transmission and reflection coefficient of the system itself is for the input signal. 
The input signal here includes both the signal that we want to transmit and the thermal noise. 
If the temperature of the environment is high, the proportion of information in the corresponding input signal will be low. In the transmission process of the system when the transmission coefficient $T=1$, the actual signal transmission coefficient will be greatly reduced due to the reduction of the signal ratio in noised input signal. 
Therefore, in order to improve the transmission rate of the signal, we need to lower the temperature. 
Conversely, if $T=0$, the system will indiscriminately isolate both the signal and the input environment noise. 
The whole system is robust to temperature. 
This robustness is certainly not unlimited, because our calculation is based on the asymmetric property of the quantum correlation of the system, so this robustness is also built on the premise of not destroying the quantum property of the system.

\section{Discussion and Conclusions}\label{sec4}
A nonlocal three body interaction is constructed in hybrid optomechanical systems.
Based on our nonlocal interaction system, we can build a nonlocal photon-phonon circulator and realize the quantum control of nonlocal optical mode to mechanical mode.
As long as we let the accumulated phase reach a specific value, that is $\theta=\pm 7 \pi/6$ or $\theta=\pm 11\pi/6 $, we can realize the nonlocal nonreciprocal transmission between the mechanical state and the optical state.
In addition, the coupling coefficient between systems can be used to control or regulate the nonreciprocity of the system. For the effective coupling coefficients $V_0$ and $G_{j0}$, we can control the nonreciprocity by controlling the adjustable parameters, that is, the laser strength $\varepsilon_{j}$ driving or detuning $\omega_{dj}$.
The corresponding relationship and controlling results are shown in the \esref{peff} and \fref{fig5}, respectively.
It is worth noting that, in the hypothesis of lossless fiber (or free space) in our scheme, we ignore the influence of environmental dissipation in the transmission process, and only consider the input noise of the optical cavity. 
If the dissipation of the transmission process is taken into account, it will inevitably introduce more noise, namely one need to add transmission noise to the expression for $\hat{A}_{in}$. 
Fortunately, this noise is also similar to the input noise of optical cavity, which are uncorrelated noises, and the effect of the two noises on dynamical evolution is consistent. 
So we can combine them together, and the dissipation coefficient and noise are certainly greater than the lossless condition.

For the feasibility of our scheme, the optomechanical system of various frequency bands has been realized in experiments.
The parameters used in our protocol can be easily implemented in typical micro- \cite{nature.494.211,nature.471.204} or nano- \cite{nature.507.81,naturephysics.4.555} optomechanical systems with $\kappa/\omega_{m} \rightarrow 0.1$.
Low loss remote optical transmission is realized in optical fiber or free space  \cite{NC.12.2381,PhysRevLett.115.093603,science.356.1140}.
Therefore, our protocol provides an executable platform for the implementation of nonlocal and nonreciprocal phonon-photon transmission or control,  and eventually provides the basis for applications on quantum information processing or quantum networking.

\section{Acknowledgments}
We thank Rui-Jie Xiao and Yang Zhang for instructive discussions.

\begin{widetext}
\section*{Appendix A: Effective Hamiltonian of system}
 According to the Hamiltonian of the system, the dynamics of the system is described by the quantum Langevin equations~(QLEs),
\begin{align}
 & \dot{\hat{a}}_1 = -[i(\Delta_{c1}+\sum_{j=1,2}g_j(\hat{b}_j^{\dag}+\hat{b}_j))+\frac{\kappa_1}{2}]\hat{a}_1-iJ\hat{a}_2-i\varepsilon_1+\sqrt{\kappa_1}\hat{a}_{in,1}, \\
 &\dot{\hat{a}}_2 = -(i\Delta_{c2}+\frac{\kappa_2}{2})\hat{a}_2-iJ\hat{a}_1-i\varepsilon_2+\sqrt{\kappa_2}\hat{a}_{in,2}, \\
&  \dot{\hat{b}}_j =-(i\omega_{mj}+\frac{\gamma_j}{2})\hat{b}_j-ig_j\hat{a}_1^{\dag}\hat{a}_1-iV\hat{b}_{3-j}+\sqrt{\gamma_j}\hat{b}_{in,j},
\end{align}
where $\Delta_{cj}=\omega_{cj}-\omega_{dj}$ is the detuning between the driving field and the cavity frequency.
After choose the suitable sideband with $\Delta_{c1}\approx \omega_{mj}$.
The linearized quantum Langevin equations of the system under the rotating-wave approximation are then governed by,
\begin{align}
 &\delta\dot{\hat{a}}_1= -(i\Delta_{c1}'+\frac{\kappa_1}{2})\delta\hat{a}_1-i\sum_{j=1,2}G_j\delta\hat{b}_j-iJ\delta\hat{a}_2+\sqrt{\kappa_1}\hat{A}_{in,1}, \label{qpart1}\\
&\delta\dot{\hat{a}}_2= -(i\Delta_{c2}+\frac{\kappa_2}{2})\delta\hat{a}_2-iJ\delta\hat{a}_1+\sqrt{\kappa_2}\hat{A}_{in,2} \label{eqa2}, \\
&\delta\dot{\hat{b}}_j=-(i\omega_{mj}+\frac{\gamma_j}{2})\delta\hat{b}_j-iG_j^*\delta\hat{a}_1-iV\delta\hat{b}_{3-j}+\sqrt{\gamma_j}\hat{b}_{in,j},\label{eqbj}
\end{align}
where $\Delta_{c1}'=\Delta_{c1}+\sum_{j=1,2}g_j(\beta_j+\beta_j^*)$.
The above differential equations of quantum part can be formally integrated as
\begin{align}
 & \delta\hat{a}_1(t)=\delta\hat{a}_1(0)e^{-(i\Delta_{c1}'+\frac{\kappa_1}{2})t}+\int^{t}_{0} d{\tau}e^{-(i\Delta_{c1}'+\frac{\kappa_1}{2})(t-{\tau})}[-i(\sum_j G_j\delta\hat{b}_j+J\delta\hat{a}_2)+\sqrt{\kappa_1}\hat{A}_{in,1}] ,\label{lz1}\\
 & \delta\hat{a}_2(t)=\delta\hat{a}_2(0)e^{-(i\Delta_{c2}+\frac{\kappa_2}{2})t}+\int^{t}_{0} d{\tau}e^{-(i\Delta_{c2}+\frac{\kappa_2}{2})(t-{\tau})}[-iJ\delta\hat{a}_1+\sqrt{\kappa_2}\hat{A}_{in,2}] ,\\
 & \delta\hat{b}_j(t)=\delta\hat{b}_j(0)e^{-(i\omega_{mj}+\frac{\gamma_j}{2})t}+\int^{t}_{0} d{\tau}e^{-(i\omega_{mj}+\frac{\gamma_j}{2})(t-{\tau})}[-i(G_j^*\delta\hat{a}_1+V\delta\hat{b}_{3-j})+\sqrt{\gamma_j}\hat{b}_{in,1}] ,
\end{align}
Under the condition $\Delta_{c2}\gg J$ and $\omega_{mj}\gg\{G_j,V\}$,
$\delta\hat{a}_2$ and $\delta\hat{b}_j$ can be approximately regarded as free evolution.
Thus, the dynamics of  $\delta\hat{a}_2$ and $\delta\hat{b}_j$ are expressed approximately as

\begin{align}\label{eqap}
& \delta\hat{a}_2(t)\approx\delta\hat{a}_2(0)e^{-(i\Delta_{c2}+\frac{\kappa_2}{2})t}+\sqrt{\kappa_2}A_{in,2}'(t),\nonumber\\
& \delta\hat{b}_j(t)\approx\delta\hat{b}_j(0)e^{-(i\omega_{mj}+\frac{\gamma_j}{2})t}+\sqrt{\gamma_j}B_{in,j}(t),
\end{align}
where $A_{in,2}'(t)=\int^{t}_{0} d{\tau}e^{-(i\Delta_{c2}+\frac{\kappa_2}{2})(t-{\tau})}\hat{a}_{in,2}$, $B_{in,j}(t)=\int^{t}_{0} d{\tau}e^{-(i\omega_{mj}+\frac{\gamma_j}{2})(t-{\tau})}\hat{b}_{in,j}$.
Substituting \esref{eqap} into the equation of $\delta\hat{a_1}(t)$, we have
\begin{align}
\delta\hat{a}_1(t)=&\delta\hat{a}_1(0)e^{-(i\Delta_{c1}'+\frac{\kappa_1}{2})t}\nonumber\\
&+\int^{t}_{0} d{\tau}e^{-(i\Delta_{c1}'+\frac{\kappa_1}{2})(t-{\tau})}
[-i\sum_j G_j\delta\hat{b}_j(0)e^{-(i\omega_{mj}+\frac{\gamma_j}{2})\tau}-iJ\delta\hat{a}_2(0)e^{-(i\Delta_{c2}+\frac{\kappa_2}{2})\tau}]+\sqrt{\kappa_1}\hat{a}_{in},
\end{align}
where $\sqrt{\kappa_1}\hat{a}_{in}=\int^{t}_{0} d{\tau}e^{-(i\Delta_{c1}'+\frac{\kappa_1}{2}](t-{\tau})}[-i\sum_j G_j\sqrt{\gamma_j}\hat{B}_{in,j}(\tau)-iJ\sqrt{\kappa_2}\hat{A}_{in,2}'(\tau)+\sqrt{\kappa_1}\hat{A}_{in,1}]$.
Under the condition $\kappa_1\gg\{\gamma_1,\gamma_2,\kappa_2\}$, $e^{-\kappa_1t}$ is a fast dissipation term, which can be ignored in dynamics.
Then we derive
\begin{align}
&\delta\hat{a}_1(t)\approx \sum_j \frac{-iG_j\delta\hat{b}_j(t)}{i(\Delta_{c1}'-\omega_{mj})+\frac{\kappa_1-\gamma_j}{2}}+\frac{-iJ\delta\hat{a_2}(t)}{i(\Delta_{c1}'-\Delta_{c2})+\frac{\kappa_1-\kappa_2}{2}}+\sqrt{\kappa_1}\hat{a}_{in},
\end{align}

Now substituting $\delta\hat{a}_1(t)$ into \eref{eqa2} and \esref{eqbj}.
We obtain the approximated expressions,
\begin{align}
&\delta\dot{\hat{a}}_2=-(i\Delta_{eff}+\frac{\kappa_{eff}}{2})\delta\hat{a}_2+i\sum_j G_j'\delta\hat{b}_j+\sqrt{\kappa_{eff}}\hat{A}_{in} ,\\
&\delta\dot{\hat{b}}_j=-(i\omega_{effj}+\frac{\gamma_{effj}}{2})\delta\hat{b}_j+iG_j''\delta\hat{a}_2+iV_j\delta\hat{b}_{3-j}+\sqrt{\gamma_{effj}}\hat{B}_{in,1},
\end{align}
the effective parameters are
\begin{eqnarray*}
\Delta_{eff}&=&\Delta_{c2}-\xi_c(\Delta_{c1}'-\Delta_{c2}),\\
\omega_{effj}&=&\omega_{mj}-\xi_{mj}(\Delta_{c1}'-\omega_{mj}),\\
\kappa_{eff}&=&\kappa_2+\xi_1(\kappa_1-\kappa_2),\\
\gamma_{effj}&=&\gamma_j+\xi_{mj}(\kappa_1-\gamma_j),\\
G_j'&=& J G_j \xi_j,\\
G_j'' &=& J G_j^* \xi,\\
V_j &=& G_j^* G_{3-j} \xi_{3-j}-V,
\end{eqnarray*}
where
\begin{eqnarray*}
\xi_c &=& \frac{J^2}{(\Delta_{c1}'-\Delta_{c2})^2+\frac{(\kappa_1-\kappa_2)^2}{4}},\\
\xi_{mj} &=& \frac{|G_j|^2}{(\Delta_{c1}'-\omega_{mj})^2+\frac{(\kappa_1-\gamma_j)^2}{4}},\\
\xi_j &=& \frac{(\Delta_{c1}'-\omega_{mj})+\frac{i(\kappa_1-\gamma_j)}{2}}{(\Delta_{c1}'-\omega_{mj})^2+\frac{(\kappa_1-\gamma_j)^2}{4}},\\
\xi &=& \frac{(\Delta_{c1}'-\Delta_{c2})+\frac{i(\kappa_1-\kappa_2)}{2}}{(\Delta_{c1}'-\Delta_{c2})^2+\frac{(\kappa_1-\kappa_2)^2}{4}},
\end{eqnarray*}
the corresponding effective noise operators are,
\begin{eqnarray*}
\sqrt{\kappa_{eff}} \hat{A}_{in} &=&-iJ\hat{a}_{in}+\sqrt{\kappa_2}\hat{A}_{in,2},\\
\sqrt{\gamma_{effj}} \hat{B}_{in,j} &=& -iG_j^*\hat{a}_{in}+\sqrt{\gamma_j}\hat{b}_{in,j},
\end{eqnarray*}
Under the condition $\Delta_{c2}=\omega_{mj}=\omega_m$, $\kappa_1 \gg \{\gamma_1,\gamma_2,\kappa_2\}$ and $\{ J,V\} \in Real$, we have $G_j'^*=G_j''$ and $V_1^*=V_2$.
The effective Hamiltonian of the system can be obtained

\begin{equation}
H_{eff}=\Delta_{eff} \delta \hat{a}_2^{\dag} \delta \hat{a}_2+\sum_{j=1}^2 \omega_{effj} \delta \hat{b}_j^{\dag} \delta \hat{b}_j-(\sum_{j=1}^2 G_j' \delta \hat{a}_2^{\dag} \delta \hat{b}_j+V_1 \delta \hat{b}_1^{\dag} \delta \hat{b}_2+h.c.).
\end{equation}
In the discussion of the main text, we are not limited to the case where the coupling coefficients are real, so our calculation is start from the effective dynamic equations rather than the effective Hamiltonian.

\section*{Appendix B: Nonlocality of our system}

According to Ref .~\cite{PhysRevLett.88.040406}, Bell-CHSH inequalities for continuous variable (CV) systems is expressed as
\begin{equation}
\mathcal{B}_{CHSH}=(\mathbf{a}\cdot \hat{s}_1)\otimes (\mathbf{b}\cdot \hat{s}_2)+(\mathbf{a}\cdot \hat{s}_1)\otimes (\mathbf{b}' \cdot \hat{s}_2)+(\mathbf{a}' \cdot \hat{s}_1)\otimes (\mathbf{b}\cdot \hat{s}_2)-(\mathbf{a}' \cdot \hat{s}_1)\otimes (\mathbf{b}' \cdot \hat{s}_2).
\end{equation}
Here $\mathbf{a}$, $\mathbf{a}'$, $\mathbf{b}$ and $\mathbf{b}'$ are four unit three-dimensional vectors.
$\hat{s}_j$ is defined as
\begin{equation}
\mathbf{a}\cdot \hat{s}=\hat{s}_z \cos \theta_a+\sin \theta_a(e^{i\phi_a}\hat{s}_{-}+e^{-i\phi_a}\hat{s}_{+}),
\end{equation}
where $\theta_a$ and $\phi_a$ are the polar and azimuthal angle of $\mathbf{a}$, respectively.
The expression of operators are as follows
\begin{eqnarray*}
\hat{s}_z &=& \sum_{n=0}^{\infty} [|2n+1 \rangle \langle 2n+1 |-|2n \rangle \langle 2n |],\\
\hat{s}_{-}&=&\sum_{n=0}^{\infty}[|2n \rangle \langle 2n+1 |=(\hat{s}_{+})^{\dag}.
\end{eqnarray*}

Then local realistic theories impose the following Bell-CHSH inequality \cite{PhysRevLett.68.3259}:
\begin{equation}
|\langle \mathcal{B}_{CHSH} \rangle| \leq 2.
\end{equation}
Thus, the nonlocality of our system will be confirmed as long as the above inequality is violated.
For the convenience of discussion, we ignore dissipation and choose the direction vectors to satisfy condition $\mathbf{a}(\mathbf{b})=\mathbf{a}'(\mathbf{b}')$.
The average value of  $\mathcal{B}_{CHSH}$  is simplified to
\begin{eqnarray}
\langle \mathcal{B}_{CHSH} (t)\rangle&=&2 \langle (\mathbf{a}\cdot \hat{s}_1(t))\otimes (\mathbf{b}\cdot \hat{s}_2(t)) \rangle \nonumber\\
&=& 2 \{ \cos \theta_a  \cos \theta_b \langle \hat{s}_{z,1} \hat{s}_{z,2} \rangle (t)+\cos \theta_a  \sin \theta_b \langle \hat{s}_{z,1} (e^{i\phi_a}\hat{s}_{-,2}+e^{-i\phi_a}\hat{s}_{+,2}) \rangle (t) \nonumber\\
&&+\sin \theta_a  \cos \theta_b \langle (e^{i\phi_a}\hat{s}_{-}+e^{-i\phi_a} \hat{s}_{+,1})\hat{s}_{z,2} \rangle (t) 
+\sin \theta_a  \sin \theta_b [e^{i(\phi_a+\phi_b)}\langle \hat{s}_{-,1} \hat{s}_{-,2} \rangle (t)\nonumber\\
&&+e^{-i(\phi_a+\phi_b)}\langle \hat{s}_{+,1} \hat{s}_{+,2} \rangle (t)+e^{i(\phi_a-\phi_b)}\langle \hat{s}_{-,1} \hat{s}_{+,2} \rangle (t)+e^{-i(\phi_a-\phi_b)}\langle \hat{s}_{+,1} \hat{s}_{-,2} \rangle (t)] \}.
\end{eqnarray}
The dynamic of operator $\mathcal{B}_{CHSH}(t)$ depends on the effective Hamiltonian of three bodies.
But Bell-CHSH inequality is to investigate the nonlocality between two bodies.
It is worth noting that, in our system, the position of $\dot{b}_j$ is equivalent, so we only need to consider one of the mechanical oscillators.
Without lose of generality, we choose subscript $1$ to represent mode $\hat{b}_1$ and $2$ to represent mode $\hat{a}_2$.
By using the time evolution operator $U=\exp(-i H_{eff}t)$, we can calculate the time evolution of the mean values of the second-order moments, $\langle \hat{s}_{i,1} \hat{s}_{j,1}\rangle, \{i,j \} \in\{z,+,-  \}$, so as to investigate $\mathcal{B}_{CHSH}(t)$.
The free term in the Hamiltonian will only add a time-dependent phase factor to the time evolution operator, and will not change the distribution of the quantum state, but the interaction term $ G_1' \delta \hat{a}_2^{\dag} \delta \hat{b}_1+h.c.$ will change the distribution of quantum states.
Since we only need to prove the existence of nonlocality here, we can discuss it with a special case.
Assuming the initial state is $| \psi\rangle=\sum_{k_1}m(k_1)|2k_1+1\rangle_1 \otimes \sum_{k_2}m(k_2)|2k_2+1 \rangle_2$, the effect of Hamiltonian on its evolution is
\begin{eqnarray*}
\delta \hat{b}_1^{\dag}\delta\hat{b}_1 |2k_1+1\rangle_1|2k_2+1 \rangle_2 & \rightarrow& |2k_1+1\rangle_1|2k_2+1 \rangle_2, \\
\delta \hat{a}_2^{\dag}\delta\hat{a}_2 |2k_1+1\rangle_1|2k_2+1 \rangle_2 & \rightarrow& | |2k_1+1\rangle_1|2k_2+1 \rangle_2, \\
\delta\hat{b}_1^{\dag}\delta\hat{a}_2  |2k_1+1\rangle_1|2k_2+1 \rangle_2 & \rightarrow& |2k_1+2 \rangle_1 |2k_2 \rangle_2, \\
\delta\hat{b}_1\delta\hat{a}_2^{\dag}  |2k_1+1\rangle_1|2k_2+1 \rangle_2 & \rightarrow& |2k_1 \rangle_1 |2k_2+2 \rangle_2.
\end{eqnarray*}
After the action of the effective Hamiltonian, a general final state in the subspace $\{|2k_0 \rangle,|2k_0 +1\rangle,|2k_0+2 \rangle \}$ can be expressed as (under the low excitation approximation in weak-driving regime, we have $k_0=0$ \cite{PhysRevA.90.023849,OE.29.36167})
\begin{equation}
|\psi\rangle_f=|\phi\rangle_1 \otimes |\phi\rangle_2,
\end{equation}
where $|\phi\rangle=m_1 |2k_0 \rangle+m_2 |2k_0 +1\rangle+m_3|2k_0+2 \rangle$, here $m_j$ is the normalized superposition coefficient, and we have $\sum_j |m_j|^2 =1 $.
Since the interaction between $\delta \hat{b}_1$ and $\delta \hat{a}_2$ is conjugate,  we can safely set $|m_1|$ and $|m_2|$ equal.
Thus, we can get the average value of $\mathcal{B}_{CHSH}$ on this final state under the condition $\phi_a=\phi_b=0$,
\begin{eqnarray}
\langle\mathcal{B}_{CHSH} \rangle &=& 2 [ \cos \theta_a \cos \theta_b+\cos \theta_a \sin \theta_b(\alpha_1^* \alpha_2+h.c.)+\sin \theta_a \cos \theta_b(\beta_1^* \beta_2+h.c.)\nonumber\\
&&+2\sin \theta_a \sin \theta_b(\alpha_1^* \alpha_2\beta_1 \beta_2^*+h.c.)],
\end{eqnarray}
where $\beta_j$ and $\alpha_j$ denotes the superposition coefficient of mode $\delta \hat{b}_1$ and $\delta \hat{a}_2$, respectively. 
We can see that, when $G_j'=0$, the coefficients $\alpha_1$ and $\beta_1$ are also equal to zero. 
At this time, the maximum value of $\langle\mathcal{B}_{CHSH} \rangle$ is $2$, which does not violate Bell-CHSH inequality, the system embodies locality.
When $G_j'\neq 0$, under the sideband condition and set $\{ \theta_a=\theta_b=\theta, \alpha_j=\beta_j\}$, the above formula can be rewritten as
\begin{eqnarray}
\langle\mathcal{B}_{CHSH} \rangle &=& 2 ( \cos \theta+\sin\theta  \alpha_2 \sqrt{1-|\alpha_2|^2})^2,
\end{eqnarray}
Through the numerical simulation in the main text, it can be confirmed that $\langle\mathcal{B}_{CHSH} \rangle$ can be greater than $2$, that is, our system can reflect nonlocality in the presence of $G_j'$.
The upper bound violated by Bell inequality is $2\sqrt{2}$. 
The designed quantum state to reach this upper bound is discussed in Ref .~\cite{PhysRevLett.88.040406}. 
In our paper, we only need an existence proof to prove that our scheme is nonlocal under the remote interaction $J$.
\end{widetext}

\bibliography{nnoc}

\end{document}